\documentclass[doublecol]{epl2}
\usepackage{epsf}

\usepackage{color}
\usepackage{graphicx}
\usepackage{dcolumn}
\usepackage{bm}
\usepackage{amssymb}
\usepackage{amsmath}

\title{Flow Generation by Rotating Colloids in Planar Microchannels}

\author{Ingo O.~G{\"o}tze\thanks{E-mail: \email{i.goetze@fz-juelich.de}} 
      \and Gerhard Gompper\thanks{E-mail: \email{g.gompper@fz-juelich.de}}}
\shortauthor{G{\"o}tze \etal}
\shorttitle{Rotating colloids in microchannels}

\institute{
  Theoretical Soft Matter and Biophysics,
  Institut f\"ur Festk\"orperforschung, Forschungszentrum J\"ulich,
  52425 J\"ulich, Germany, EU\\
}
\pacs{47.61.-k}{Micro- and nano- scale flow phenomena}
\pacs{83.10.Rs}{Computer simulation of molecular and particle dynamics}
\pacs{82.70.Dd}{Colloids}

\abstract{
Non-equilibrium structure formation and conversion of 
spinning to translational motion 
of magnetic colloids driven by an external rotating magnetic     
field in microchannels is studied by
particle-based mesoscale hydrodynamics simulations.
For straight channels, laning is found.
In ring channels, the channel curvature breaks symmetry and
leads to a net fluid transport 
around the annulus with the same rotational direction
as the colloidal spinning direction.
The dependence of the translational velocity 
on channel width, ring radius, colloid concentration, 
and thermal motion is predicted.
}

\begin{document}

\maketitle

Externally actuated and self-propelled micro- and nano-rotators
show an intriguing variety of non-equilibrium structure formation and
dynamics. Examples of such systems include super-paramagnetic
colloidal particles in a rotating magnetic field
\cite{grzy00,grzy02a,bech06,kavc09},
dipolar colloids in a rotating electric field~\cite{elsn09},
colloidal dimers rotated by laser tweezers \cite{terr02}
and biological swimmers such as Volvox algae~\cite{dres09}.
The collective behavior of rotator suspensions is governed by  
hydrodynamic interactions.
While a single rotator cannot show any directed translational
motion, in ensembles of several rotators their spinning motion   
leads to an enhanced translational diffusion~\cite{llop08}.
Other interesting phenomena in suspensions of rotators are
stable bound states of spinning Volvox algae~\cite{dres09},
spontaneous pattern formation of spinning magnetic disks 
at liquid-air interfaces in the form of rotating
hexagonal crystal structures~\cite{grzy00},
and spinning colloids, placed asymmetrically in a microfluidic
channel, which act as micropumps~\cite{terr02,bech06,kavc09}.

A key prerequisite for the conversion of rotational into translational 
motion is symmetry breaking in confinement.
For a single rotating cylinder close to a planar wall (and oriented 
parallel to the wall), 
the resulting hydrodynamic forces vanish since the viscous stress is 
balanced by the 
dynamic pressure field~\cite{jeffrey81}; thus, a single wall 
is not sufficient to generate translational motion
of a cylinder (in contrast to a sphere, which experiences
a force parallel to the wall in the direction expected for a rolling 
motion~\cite{goldman67}).
However, in the presence of a second parallel wall, 
more fluid is pumped through the wide gap between cylinder and wall 
than through the narrow gap, causing a net flow in the channel and 
a reaction force parallel to the wall on the cylinder --- unless 
it is centered between the walls \cite{sen96,DeCour98}.
Pumps based on this principle, with a typical channel diameter 
in the centimeter range and a 
fixed position of the rotating cylinder, have been studied both 
experimentally and numerically~\cite{sen96,DeCour98}.
The goal of microfluidics is to work with much smaller length
scales, on the order of micrometers or smaller. Indeed, micron-sized 
pumps employing rotating colloids have been contructed recently 
\cite{terr02,bech06,kavc09}.

With further miniaturization of microfluidic systems,
thermal fluctuations become increasingly important. Furthermore,
a sufficiently large concentration of particles is required to achieve 
reasonable flow velocities. Thus, we investigate a system of 
spinning colloidal particles, 
which are free to move in a channel, are small enough to display
thermal Brownian motion, and have a finite concentration. 
Pairs of co-rotating cylinders in an unbounded viscous fluid are
known to mutually exert forces on each other that are perpendicular 
to their connecting line~\cite{ueda03}, 
which implies a circling motion. At higher
concentrations, coordinated motion occurs~\cite{llop08}.
 
\begin{figure}
\begin{center}
\vspace{0.5cm}
\includegraphics[width=7.5cm]{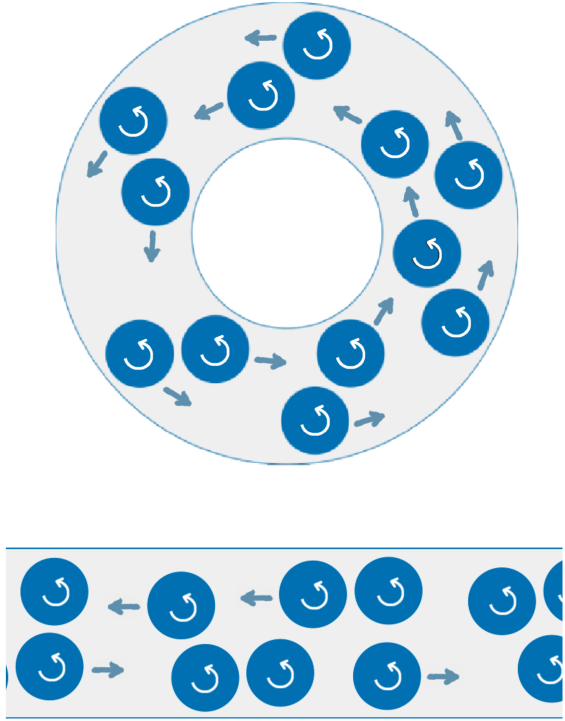}
\caption{\label{fig:snapshot}
(Color online)
Simulation snapshots of rotating colloids in a ring channel 
(upper part) with median radius $R=2.4\sigma$ and width 
$D=2.0\sigma$, where $\sigma$ is the colloid diameter, and of a
section of a linear channel (lower part) with $D=2.5\sigma$.
The rotational and translational motion of the colloids
is illustrated by the arrows.
For movies, see also footnote 1.
}
\end{center}
\end{figure}

In a channel, many-body effects and the presence of the walls 
modify the interactions, and the situation becomes much more complex.
Symmetry implies that there cannot be any net fluid or colloid 
transport in a straight channel.
However, we find that the particle distribution across the channel 
causes {\it local} directed motion and leads to lane formation.
In ring channels, the additional symmetry breaking leads to a 
net translational motion along the channel,
as illustrated in Fig.~\ref{fig:snapshot}.
Simulation animations of spinning colloids in 
a linear channel ({\tt linear\_channel.avi}) in a
circular channel ({\tt ring\_channel.avi}) are available in the online 
supporting material.
\footnote{
   The movie {\tt ring\_channel.avi} shows a simulation 
   animation of 2D colloids in a ring channel of median radius 
   $R=2.4 \sigma$ and width $D=2.0 \sigma$. The resulting fluid 
   velocity field is displayed by arrows. The movie 
   {\tt linear\_channel.avi} shows counter-clockwise 
   rotating 2D colloids in a linear channel of width $D=2.5 \sigma$.
   In the displayed system, the colloid area fraction is $\Phi=0.38$.
   In both movies, a constant external torque $L_{\rm ext}=150 k_BT$ 
   is applied.
}
We address the fundamental question how symmetry breaking due to
channel curvature 
generates net transport in the presence of thermal fluctuations.
For simplicity, we restrict ourselves to two-dimensional systems.

The fluid is modeled by multi-particle collision dynamics (MPC), a 
mesoscale hydrodynamics simulation technique that naturally includes 
thermal fluctuations~\cite{male99,kapr08,gg:gomp09a}.
Here, the fluid is represented by point particles; their dynamics
proceeds in two alternating steps.  In the streaming step, the fluid 
particles move ballistically.  Subsequently, they are sorted into 
the cells of a
square lattice with lattice constant $a$. New relative velocities
(with respect to the center-of-mass velocity of each cell) 
are assigned to all fluid particles in a collision cell, which mimics 
the simultaneous interaction of these particles.
The algorithm is constructed such that linear momentum is conserved 
in each cell, in order to obtain correct hydrodynamic behavior.
For a study of hydrodynamics of rotating colloids with fixed torque, 
it is essential to employ a variant of MPC which also conserves 
angular momentum~\cite{gg:gomp07b,gg:gomp07h,comm_MPC}.

\begin{figure}
\begin{center}
\includegraphics[width=8cm]{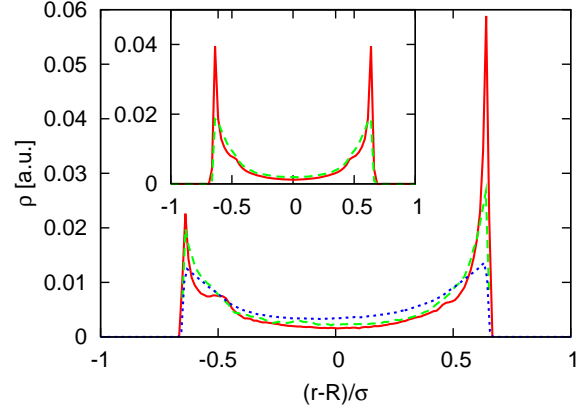}
\caption{\label{fig:density}
(Color online)
Colloid density distribution across a ring channel of 
median radius $R=2.4 \sigma$ and width $D=2.5 \sigma$, 
with area fraction $\Phi=0.25$. Data are shown for torques
$L_{{\rm ext},0}$ (full red line) and $L_{{\rm ext},0}/3$ 
(dashed green line), as well as for colloids at thermal 
equilibrium (dotted blue line).
Inset: Same for a straight channel with torques $L_{{\rm ext},0}$ 
(full red line) and $L_{{\rm ext},0}/9$ (dashed green line). }
\end{center}
\end{figure}

The colloids are modeled as discs of diameter $\sigma$ with no-slip 
boundary conditions.  For the colloid-colloid and the colloid-wall 
interactions, a purely repulsive (shifted truncated) Lennard-Jones (LJ) 
potential is added. Its range 
 $\delta_{\rm LJ} \approx 0.1 \sigma$ is chosen to be small compared
to the colloid radius, but large enough 
to ensure that there is at least one collision 
box of fluid between surfaces. This guarantees that hydrodynamic 
interactions are captured quantitatively even at the smallest
colloid distances \cite{gg:gomp07h}.
We apply a constant external torque $L_{\rm ext}$ to all colloids, 
causing the colloids to spin counter-clockwise about their axes with 
an average angular velocity 
${\Omega}_\infty=L_{\rm ext}/(\eta\pi\sigma^2)$ in the bulk,
where $\eta$ is the viscosity of the fluid.
In a channel, the colloids develop a characteristic translational  
velocity $v_0$.
The P\'eclet number for the rotating colloids is then defined by 
${\rm Pe}=\sigma v_0/D_t$, where $D_t$ is the 
(geometry-dependent) translational diffusion 
coefficient. The P\'eclet number depends both on the channel
geometry and the external torque. For the largest torque considered,
$L_{{\rm ext},0}=150 k_BT$,  the P\'eclet number varies 
between ${\rm Pe}={\cal O}(10)$ (narrowest channel) and 
${\rm Pe}={\cal O}(10^2)$ (widest channel). For the widest channel, 
the P\'eclet number varies 
linearly with the colloid torque, with ${\rm Pe}={\cal O}(10)$ 
for the smallest torque $L_{{\rm ext},0}/9=16.7 k_BT$.  
If not mentioned otherwise, the torque $L_{{\rm ext},0}$ is employed.  

First, we investigate straight channels that are wide enough to 
allow colloids to pass each other, with widths $D=2.5 \sigma$ and 
$D=3 \sigma$. 
We observe that the spinning colloids 
move in opposite directions at opposing walls
(see movie {\tt linear\_channel.avi}),   
and with increasing concentration approach the walls very closely, 
see Fig.~\ref{fig:density}, with 
a distance only slightly larger than $\sigma/2 + \delta_{\rm LJ}$.
A comparison with simulations for smaller spinning 
velocities, see Fig.~\ref{fig:density}, shows that the peaks in 
the density profile broaden with decreasing $L_{\rm ext}$. 
A weak density increase near the walls due to the
finite colloid concentration remains even in thermal equilibrium.
For wider channels, the surface excess is less pronounced.

\begin{figure}
\begin{center}
\includegraphics[width=8cm]{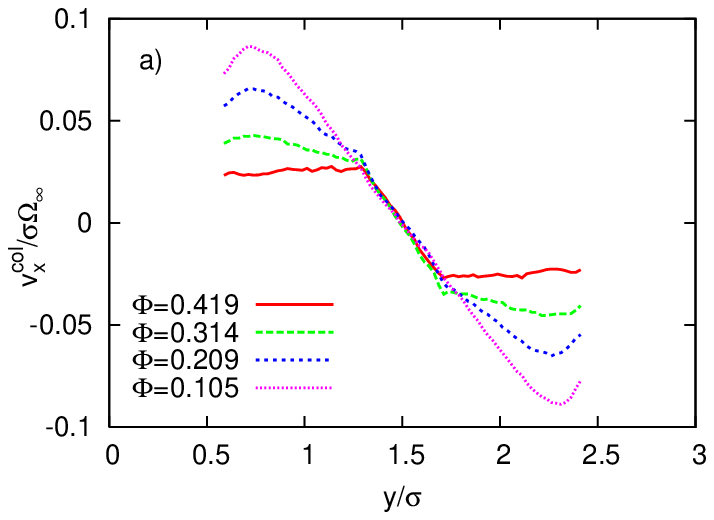}
\includegraphics[width=8cm]{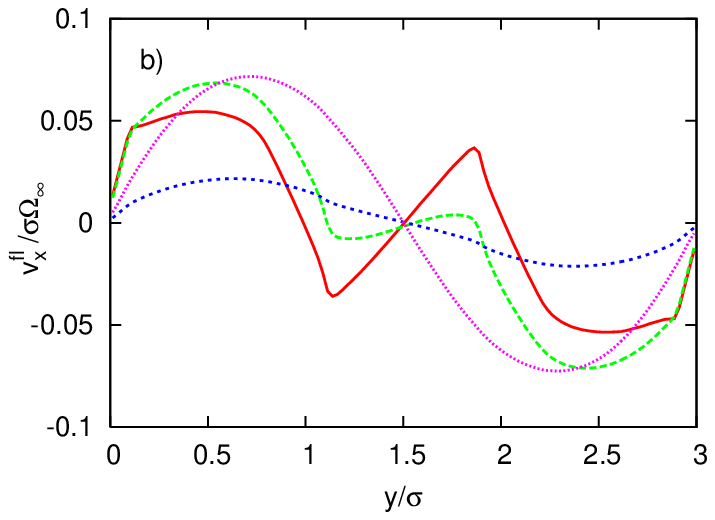}
\caption{\label{fig:lc_vp}
(Color online)
(a) Colloid velocity profiles for various colloid 
area fractions $\Phi$ for a straight channel of width $D=3 \sigma$. 
(b) The corresponding fluid velocity profiles.
}
\end{center}
\end{figure}

On a qualitative level, this behavior can be understood as follows.
When two spinning colloids pass each other in the channel, 
the hydrodynamic forces mutually exerted on each other enhance 
this directed motion. For symmetry reasons, colloids spinning in 
the center of the channel cannot propel themselves forward; however,
they also cause a local fluid motion, and hence push 
nearby colloids towards one of the walls, while simultaneously being 
pushed to the opposite wall.

Colloid velocity 
profiles across the channel are shown in Fig.~\ref{fig:lc_vp}a
for $D=3 \sigma$ and various colloid densities, together with 
the corresponding fluid velocity profiles in Fig.~\ref{fig:lc_vp}b.
The fluid velocity profile is a result of the combination of 
colloid motion pushing the fluid column ahead of it forward,  
the surface velocity due to colloid rotation, the no-slip
boundary conditions on the walls, and 
the velocity field generated by the colloids near the opposite wall.
Moreover, since a constant torque is exerted on the colloids, 
they spin faster at larger distances from the wall and at 
smaller (local) colloid densities. Note that the directed motion 
near the walls does {\it not} imply that individual colloids always
move in the same direction; instead, they change lanes randomly
due to thermal fluctuations and 
hydrodynamic interactions (which promote a circling motion
of near-by colloids).

\begin{figure}
\begin{center}
\includegraphics[width=8cm]{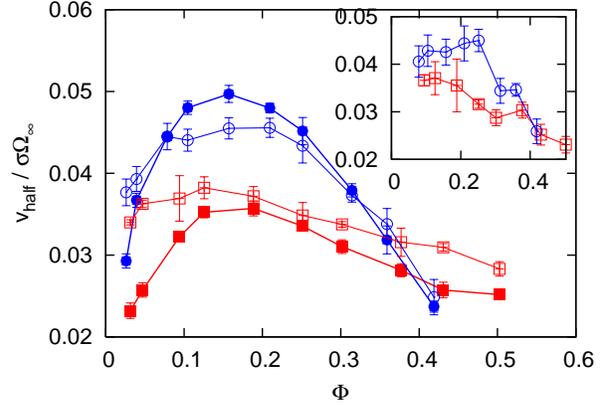}
\caption{\label{fig:lc_vhalf}
(Color online)
Average colloid velocity per half-channel as a function of the 
colloid area fraction $\Phi$ for straight channels 
of width $D=2.5 \sigma$ ($\square$, $\blacksquare$, red) and $D=3 \sigma$ 
($\circ$, $\bullet$, blue). Data are shown for three different torques,
$L_{{\rm ext},0}$ ($\blacksquare$, $\bullet$), $L_{{\rm ext},0}/3$ 
($\square$, $\circ$), and $L_{{\rm ext},0}/9$ (inset).
}
\end{center}
\end{figure}

\begin{figure}
\begin{center}
\includegraphics[width=8cm]{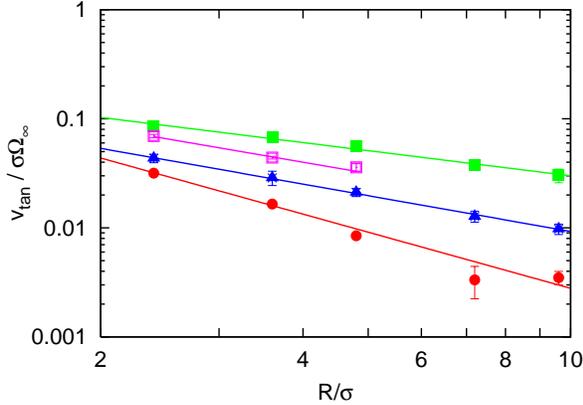}
\caption{\label{fig:vtanR}
(Color online)
Average tangential velocity $v_{\rm tan}$ of all colloids as 
a function of the median annulus radius $R$ for curved channels with 
widths $D=1.5 \sigma$ ($\bullet$, red), $2.0 \sigma$ 
($\blacksquare$, green), $2.5 \sigma$ ($\blacktriangle$, blue) and 
$3.0 \sigma$ ($\square$, magenta). 
The torque on each colloid is $L_{{\rm ext},0}$.  
}
\end{center}
\end{figure}

Although the fluid and colloid velocities averaged over the whole
channel vanish due to symmetry reasons, 
there is a directed motion in both half-channels.
To quantify this effect, we calculate the average colloid velocity 
per half-channel, as shown in Fig.~\ref{fig:lc_vhalf}.
In the limit of vanishing colloid concentration, the only forces
acting on the colloid originate from the hydrodynamic interactions 
with the walls.
In contrast to Fig.~\ref{fig:density},
the density distribution across the channel for 
a single colloid is a uniform distribution for small Pe;  
because there is no propulsion near the channel center, but a propulsion 
in opposite directions near opposing walls, the velocity per 
half-channel approaches a finite but small value in this limit 
(see Fig.~\ref{fig:lc_vhalf}).
With increasing colloid density, pairwise interactions between colloids
begin to contribute. As mentioned above, when two colloids come close 
to each other, 
they circle around each other, which implies an increased and reduced
density near the walls and in the center, respectively. 
Because the propulsion is most efficient near the walls, the 
colloid velocity in the half-channels increases when the colloids 
pass each other, and the fluid volume each colloid has to move 
decreases as $1/\Phi$, the half-channel velocity increases
with $\Phi$ at low densities.

At very high colloid concentrations, the average colloid 
velocity also becomes small because lubrication forces between neighboring
colloids and between colloids and the walls slow down the rotational
motion.
Hence, the half-channel velocity exhibits a maximum velocity at a 
finite colloid area fraction $\Phi\simeq 0.15$, see Fig.~\ref{fig:lc_vhalf}.
Moreover, the velocity is higher in the broader channel, where 
the colloids can pass each other more easily.

Fig.~\ref{fig:lc_vhalf} also demonstrates the effect of a variation
of the colloidal spinning velocity by a change of the external
torque. We observe that $v_{\rm half}/(\sigma \Omega_\infty)$ is nearly
independent of $L_{\rm ext}$, down to P\'eclet numbers as small as 
${\rm Pe}=4.5$.  
In this case, diffusive motion is clearly visible,
see movie ({\tt ring\_lowerPE.avi}) in the Supporting Material
\footnote{
   The movie {\tt ring\_lowerPE.avi} shows a simulation animation of 
   2D colloids in a ring channel of median radius $R=2.4 \sigma$ and 
   width $D=2.5 \sigma$.
   A constant external torque $L_{\rm ext}=16.7 k_BT$ is applied, a
   factor 9 smaller than for the movie {\tt ring\_channel.avi}.
}. 
This implies that the directed two-lane traffic is a very robust 
phenomenon, which is sustainable even in the presence of large thermal noise.

Fig.~\ref{fig:lc_vhalf} indicates that the colloid velocity in the
zero-density limit is lower for the highest P{\'e}clet number than for the
intermediate P{\'e}clet number. We attribute this behavior to the 
small but finite rotational Reynolds number 
Re$_{\rm rot} = \rho\Omega_\infty \sigma^2/(4\eta)$, which increases with 
increasing torque (where $\rho$ is the fluid density)
and reaches Re$_{\rm rot}=0.37$ for the largest torque; 
this inertia effect implies a small lift force (in analogy with 
Refs.~\cite{scho89,grzy00}) and 
a reduction of the colloidal density near the walls, in agreement 
with the simulated density profiles. 

In order to obtain a net transport, symmetry breaking is necessary.
The curvature of a ring channel fulfills this requirement.
We characterize the geometry of the annulus by its median 
radius $R$ and channel width $D$.
We measure the average tangential velocity $v_{\rm tan}$ of
{\em all} colloids in the 
channel for various $R$ and $D$, keeping the line density 
$N_{\rm c}\sigma/(2\pi R)$ fixed, where $N_{\rm c}$ is the 
number of colloids. Results are shown in
Fig.~\ref{fig:vtanR} for $D=1.5 \sigma$, $D=2 \sigma$ and 
$D=2.5 \sigma$ as a function of the median channel radius $R$.
We observe a total counter-clockwise tangential motion 
in the annulus that 
decreases monotonically with increasing annular radius for all 
investigated channel widths. Over the accessible range of 
annular radii, the tangential velocity $v_{\rm tan}$ is well described 
by a power-law decay, 
\begin{equation}
v_{\rm tan} \sim R^{-\gamma}.
\end{equation}
Due to the complex interplay of geometric and hydrodynamic 
effects, $v_{\rm tan}(R)$ does not obey a unique power law.
For the smallest channel width, $D=1.5$, we find  
$\gamma\simeq 1.7$, for $D=2.0\sigma$ an exponent $\gamma\simeq 0.76$,
and for wider channels, $D=2.5\sigma$ and $D=3.0\sigma$, where the colloids 
can pass each other, $\gamma=1$. This indicates an universal 
exponent for $D \ge 2.5\sigma$.

\begin{figure}
\begin{center}
\includegraphics[width=8cm]{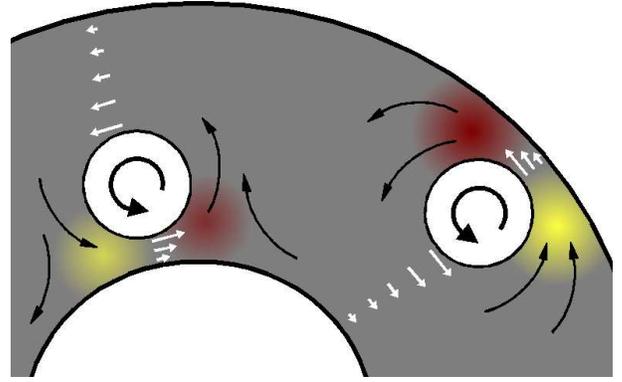}
\caption{\label{fig:curvature}
(Color online) 
Schematic illustration of colloids spinning close to the 
inner and outer confining cylinders of a ring channel.
Fluid is pumped towards or away from the walls (curved black arrows), 
giving rise to dynamic pressure.
Excess pressure is indicated in yellow (light gray), low pressure in red (dark gray).
The small straight white arrows depict the shear flow in the gaps.
}
\end{center}
\end{figure}

\begin{figure}
\begin{center}
\includegraphics[width=8cm]{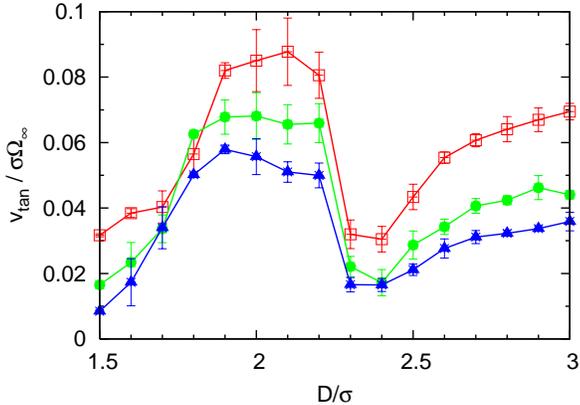}
\caption{\label{fig:vtanD}
(Color online)
Average tangential velocity $v_{\rm tan}$ of the colloids as 
a function of channel width $D$ for fixed annulus radii 
$R=2.4 \sigma$ ($\square$, red), $R=3.6 \sigma$ 
($\bullet$, green) and $R=4.8 \sigma$ ($\triangle$, blue)
for a constant line density $N_c \sigma / 2 \pi R=0.8$.
The torque on each colloid is $L_{{\rm ext},0}$. 
}
\end{center}
\end{figure}

We propose that the mechanism of net colloid transport in a ring
channel and its direction can be understood as follows.
Consider a single colloid spinning counter-clockwise about a fixed axis 
close to one of the confining walls (see Fig.~\ref{fig:curvature}).
There are three relevant contributions to the force acting on the 
colloid:
(i) The surface motion of the colloid causes a high velocity gradient 
in the small gap;
due to the corresponding shear stress, a force is exerted on the 
colloid parallel to the wall in the direction expected for a rolling
motion.
(ii) A smaller velocity gradient is induced in the wide gap,
causing a smaller force in the opposite direction.
(iii) Fluid is pumped into or out from the small gap;
the resulting pressure difference on both sides of the colloid drives 
it in the direction opposite to the direction expected for a rolling 
motion \cite{muel42,gg:gomp07h}.

For a single planar wall, these three contributions cancel exactly.
In case of a linear channel, the contributions (ii) and (iii) dominate 
over (i), causing the two way traffic.
However, in a ring channel, due to the wall curvature the dynamic pressure
changes compared to the linear channel.
Near a convex wall, the flow caused by the spinning colloid can avoid 
the wall more easily than near a concave wall.
Thus, the pressure gradient near the inner wall is smaller than near 
the outer one.
Therefore, the total force acting 
on a colloid near the outer wall is enhanced, and the force exerted on 
a colloid near the inner wall is reduced.
This explains the difference in the absolute value of the velocity near 
the confining walls that gives rise to the observed net transport in
the counter-clockwise direction.

Moreover, the circumference of the outer wall is larger than of the 
inner wall, so that
--- due to geometric reasons --- there are less colloids close to the 
inner wall that could diminish the average velocity.
The latter effect becomes more important at large $D$ and small $R$,
where the ratio of the circumferences of the confining walls increases.

In a ring channel, colloidal transport can also be achieved 
in very narrow channels,
where the colloids cannot pass each other, since the forces acting on
colloids near opposite walls are opposite in direction, but different 
in magnitude.

When the channel width $D$ is varied for a fixed median radius, 
the tangential velocity shows an interesting, non-monotonic
behavior, see Fig.~\ref{fig:vtanD}.
For narrow channels, the average volocity increases with increasing $D$ 
and reaches a maximum at $D \approx 2.1 \sigma$, where the colloids just
cannot pass each other.
With further increasing channel width, the average velocity drops to a
minimum at $D \approx 2.4 \sigma$, and then rises again.

In the narrowest channels, with $D=1.5\sigma$, we observe a low average 
tangential velocity of the colloids.
One reason for this is the strong confinement 
that leads to a reduced spinning frequency compared to wider channels.
Moreover, the colloids tend to stay close to the outer wall.
This configuration allows an unhindered circulation of the 
fluid around the inner confining cylinder in opposite direction 
(compared to the motion of the colloids). 

A maximum in the tangential velocity is reached in channels 
with $D \approx 2\sigma$ for all investigated median radii 
($2.4\sigma<R<9.6\sigma$).
At the maximum, because of the additional LJ potential, two colloid 
just cannot pass each other. Since the forces acting on 
colloids close to the inner and outer walls  
are in opposite directions, but different in magnitude,
pairs of colloids become wedged together into the 
channel (see snapshot in Fig.~\ref{fig:snapshot}) and perceive 
a net force that drives these pairs in counter-clockwise 
direction.
As they cannot pass each other, all colloids move 
in the same direction.
In this way, the colloids are forming plugs that the fluid 
can hardly pass, the fluid is efficiently dragged with 
these pairs, and a strong backflow of the fluid that would 
counteract the motion of the colloids is prevented.
In the range of $D=2.0\sigma$ to $D=2.2 \sigma$, 
we observe for $R=2.4 \sigma$ the formation of colloid triplets 
as well as pairs and combination of both (see movie 
{\tt ring\_channel.avi}), depending on the initial conditions,
where both types of clusters are stable for 
roughly 500 full circulations around the annulus.
Since all involved interactions are repulsive, the 
stabilization of the clusters is purely hydrodynamics.
The highest tangential velocity is observed for triplets.
For $R=3.6 \sigma$ and $R=4.8 \sigma$,
triplets can be observed transitionally, 
but pairs of colloids are the preferred configuration.
 
In wider channels, with $D > 2.2 \sigma$, where the colloids
can pass each other (see movie {\tt ring\_channel.avi}), 
colloids at the inner and outer walls
move in opposite directions, similar to the two-way traffic 
in the straight channels discussed above, and the tangential 
velocity drops to a minimum.
However, due to a difference in the absolute values of the colloid 
velocities at the opposite walls, the two-way traffic is superposed 
with an overall translational motion and the minimum net velocity 
is non-zero.
With further increasing channel width, the tangential 
velocity rises again.
This observation is in accordance with the case of straight 
channels, where the average velocity per half-channel is 
higher for broader channels.  
Moreover, the ratio of the circumferences of the two confining 
cylinders increases with increasing channel width, so that
there are more colloids moving in counter-clockwise direction.
Because the {\it line} density is kept constant, the {\it area}
density of the coloids will decrease with increasing $D$, implying 
that the density peaks at the walls decrease.
As the transport is mainly generated near the walls, we expect the 
velocity to decrease again with further increasing $D$.
However, the inner confining cylinder vanishes for $D=2R$ 
and the annulus loses its channel-like character well before.
In the limit $D=2R$, the ring becomes a circular cavity; in this
case, colloids are known to move in counter-clockwise direction
tangential to the wall, and experience a small lift force 
\cite{gg:gomp07h}.

Our results in Fig.~\ref{fig:vtanD} can be used to estimate typical 
transport velocities of colloids in a ring channel. For the
experimental system of Ref.~\cite{bech06}, $\Omega=125$ Hz and 
$\sigma=10 \mu$m,
we find tangential velocities $v_{\rm tan}$ in the range from
$10 \mu$m\,s$^{-1}$ to $150 \mu$m\,s$^{-1}$. This is more than an
order of magnitude larger than the transport velocities observed 
for a different channel geometry in Ref.~\cite{bech06}.

In summary, our simulations show that spinning colloids show
a rich variety of non-equilibrium states in microchannels,
from lane formation in straight channels to net transport 
in ring channels. In more complex channel networks, the fluid
flow could be manipulated very easily by controlling the positions
of some colloidal particles by laser tweezers, and to use them
as valves or to change boundary conditions. For example, net
transport in a straight channel could be induced by fixing 
colloidal particles on one side of the channel in regular 
intervals. Indeed, different kinds of boundary conditions
are another interesting possibility to break the symmetry and induce
flow, as has been pointed out recently for electrohydraulic 
pumping of dipolar molecular fluids (like water) in rotating 
electrical fields~\cite{bont09}. The manipulation of dipolar colloids
by rotating electrical fields \cite{elsn09} is another interesting 
possibility, and our results should carry over to such systems 
to a large extent.

\acknowledgments
Stimulating discussions with C.~Bechinger (Stuttgart) are gratefully
acknowledged.


\begin{thebibliography}{10}
\bibitem{comm_MPC}
In our MPC simulations, we employ the following parameters: density $\rho= 10
  a^{-2}$, collision box size $a=\sigma/10$, time step $\Delta t = 0.02
  a\sqrt{m/k_BT}$, and fluid particle mass $m$. With these parameters,
  low-Reynolds-number flow is achieved in the microchannels considered here
  \cite{gg:gomp04e}.
\end{thebibliography}

\end{document}